# Dynamic Modeling and Adaptive Controlling in GPS-Intelligent Buoy (GIB) Systems Based on Neural-Fuzzy Networks

Dangquan Zhang [1], Muhammad Aqeel Ashraf [1,2], Zhenling Liu [3], Wan-Xi Peng [1], Mohammad Javad Golkar [4], Amir Mosavi [5,6]

[1] School of Forestry, Henan Agricultural University, Zhengzhou 450002, China

[2] Department of Geology Faculty of Science, University of Malaya, 50603 Kuala Lumpur, Malaysia

[3] School of Management, Henan University of Technology, Zhengzhou 450001, China

[4] Department of Electrical and Computer Engineering, Zahedan Branch, Islamic Azad University, Zahedan, Iran

[5] Kalman Kando Faculty of Electrical Engineering, Obuda University, 1034 Budapest, Hungary

[6] School of the Built Environment, Oxford Brookes University, Oxford OX3 0BP, UK

**Abstract**

Recently, various relations and criteria have been presented to establish a proper relationship between control systems and control Global Positioning System (GPS)-intelligent buoy system. Given the importance of controlling the position of buoys and the construction of intelligent systems, in this paper, dynamic system modeling is applied to position marine buoys through the improved neural network with a backstepping technique. This study aims at developing a novel controller based on adaptive fuzzy neural network to optimally track the dynamically positioned vehicle on water with unavailable velocities and unidentified control parameters. In order to model the network with the proposed technique, uncertainties and the unwanted disturbances are studied in the neural network. The presented study aims at developing a neural controlling which applies the vectorial back-stepping technique to the surface ships, which have been dynamically positioned with undetermined disturbances and ambivalences. Moreover, the objective function is to minimize the output error for the neural network (NN) based on closed-loop system. The most important feature of the proposed model for the positioning buoys is its independence from comparative knowledge or information on the dynamics and the unwanted disturbances of ships. The numerical and obtained consequences demonstrate that the controller system can adjust the routes and the position of the buoys to the desired objective with relatively few position errors.

**1. Introduction**

Buoys have extensively been employed in different fields such as transportation, military, research and marine explorations. The development of the design and construction of buoys has been simultaneous with the growth in using buoys; therefore, many studies have been ~~are~~ performed in this field. The use of buoys in various fields i.e. transportation, military, research and marine explorations have been expanded thus far. The design and construction of buoys have been developed along with the expansion in using buoys. Besides, many studies have are presented in this area. Based on the needs today, designing and building intelligent buoys with better control and maneuver appear to be necessary. Investments in this sector by different countries clearly prove this statement. In 1997, competitions known as Association for Unmanned Vehicle (AUV) were held by the Research Institute of Naval ONR to develop and upgrade the submarines. In this competition, engineers designed systems to participate in maneuvers in the underwater environment. This competition, which was held under the name of Association for Unmanned Vehicle Systems International, linked young engineers to institutions active in the AUV communication technologies. It is worth mentioning that the active power filter (APF) which has been regarded as a self-regulating control device has the capability to discover harmonics and, in order to enhance the quality of power, inject compensation current into the grid. Besides, by applying the new intelligent concept, we can obtained better tracking control effects by control algorithms [1]. Smart Dynamic Positioning (DP) has created many probable actions which were not previously possible. The advantages of smart dynamic positioning include easy movement in naval exercises, easy positioning at the depth of water, and avoiding destructive sea beds. With the development of electrical equipment and the exploitation of the oceans to the farther and deeper seas, dynamic positioning techniques are currently applied in many applications i.e. sea exploration, intubation and drilling [2]. In [1], in order to track an APE perfect current with a restricted time-control execution, a controller based on neural network and fractional order is developed. In [3], in order to reach a better execution of APF, a fractional-order sliding mode control (FSMC) scheme, in which Recurrent Neural Network (RNN) estimator was used, has been introduced. The suggested RNN-FSMC scheme aims at combining a FSMC method with that of the RNN. In order to bring more superior control, it is in the FSMC that one can find better and adjustable rate of freedom in comparison to the integer order of FSMC. In this regard, the RNN has been applied to approach the APF nonlinear pattern. It is in [4] that the design of a NN with a double hidden layer RNN (DHLRNN) arrangement has been presented. Besides, an adaptive FSMC, which has been based on the DHLRNN, has also been suggested for a category of dynamic systems. In order to assemble the Gaussian function central vector in the DHLRNN structure and the base width, the mechanisms of adaptive adjustment and the theoretical guidance have been set up. In this sense, the stabilization of 6 sets of parameters would be reached as it is to the best values and various inputs. The network's generalization ability and accuracy would be improved by the novel DHLRNN which reduces the network weights number and speeds up the training speed of the network. Due to the fact that the input layer neurons are capable of receiving signals, they will be able to own associative memory and hasty method convergence. Consequently, superior estimate and a better dynamic ability will be achieved. In [5], a controller based on fuzzy theory and sliding mode approach has been suggested for a single-phase Photovoltaics grid-connected inverter. In this sense, it is worth mentioning that there are some uncertainties which have been the result of the inverter section variables and also some modifies in the climatic environment. Consequently, they would significantly influence the control act of the inverter. As a result, in order to roughly calculate the disturbances in real time, a disturbance observer would be designed. Besides, to modify the control system efficiency, a fuzzy system based on the upper bound estimation and observation error is applied. In [1–6], the ship's dynamic positioning controller scheme was developed with the supposition that the motion formulations could be considered as a set of fixed rotating angles. In 1990, nonlinear control theories were employed to design DP controller where linear theories had been removed. In [7], non-linear output control based on global uniform asymptotic stability were provided after observations. On the other hand, environmental problems caused by wind, waves and sea currents are imposed on the control system and affect maneuvering the dynamic positioning of ships. These problems are not negligible and are less likely to have been focused on the above-mentioned papers. In [8], a controller based on fuzzy output feedback is presented for the dynamic positioning (DP) – which exists when the unmeasured states of ships are present. It is worth noting that this is the first report on the mentioned conditions. It is with the help of a vectorial back-stepping method by which we witness the development of a control system by means of high-gain observer. It is in [9] that an adaptive robust fin controller – which has been founded on a feedforward neural network – has been suggested in order to lower the ship roll motion. Besides, they can be referred to as the environmental disturbance and the modeling errors which have been made by waves. Based on the passive nonlinear observation in [10], an output of proportional–derivative feedback control law, which has been presented for the dynamic positioning of the ship, uses position measurement. The effects of the proposed model were stated during the tests on a model of the ship into the sea at a scale of 1:70 in a laboratory. When slow unknown changes of environmental problems were considered, a nonlinear adaptive proportional–integral–derivative (PID) controller with the overall stability of the uniform DP system integrator of ships is considered [11]. In [12], a controller was created by means of combining the dynamic surface control technique by a back-stepping one. The proposed model reduces the trouble of "explosion of terms" intrinsic in the back-stepping approach, despite the fact that its implementation is simpler. In [13], two accurate linear controllers and three different comparative nonlinear model by Lyapunov's direct method were introduced to the dynamic positioning of underwater robots. These are not numerical simulations only. Rather, the laboratory results were presented, and a common position error and an initial speed were also adopted in order to compare the performance of the family of this controller, quantitatively. In recent years, a variety of intelligent methods i.e. fuzzy technique [13], NN controller [14] and predictive control vector machine-based model [15] have been employed for the DP of ships.

In [16], an interval type-2 (IT2) fuzzy method was applied and the fault detection problem was studied for continuous-time fuzzy semi-Markov jump systems (FSMJSs). Firstly, the IT2 fuzzy approach is applied in order to address the parameter uncertainty and also the FSMJSs form is considered. In this sense, the characteristic of sensor saturation would be seriously noticed in the control system. Secondly, after the construction of the IT2 fuzzy FSMJSs filter, it is used in order to deal by means of the fault finding problem. In [17], the bipartite control problem for the nonlinear multi-agent methods with input quantization over a signed digraph has been studied. Providing an appropriate distributed protocol can be regarded as the aim of the design in which the followers converge to a convex hull which contains each leader trajectory and also its opposite trajectory different in sign. Moreover, in order to estimate unmeasurable states, one needs to construct an observer. In [18], in order to design complex oscillators – i.e. free-piston Stirling engines (FPSEs) –a novel averaging-based Lyapunov approach was introduced. In this sense, the FPSE which is a class of thermal oscillators is introduced. Afterwards, the proposed technique is applied to investigate the limit cycle existence in the oscillator dynamic response as it is a required criterion for steady operation. In [19], it is the tracking control problem of an underactuated ship that is investigated. It is worth mentioning that since the method has no relative degree, the application of standard dynamic inversion is prevented. In [20], the visual DP and the visual feedback models for the ships have been presented a camera in the ship extracts the necessary information concerning the items in a 3D world. All literatures on the DP problem assume some unrealistic assumptions: that the variables are not changed during operation and that the WF model parameters are known a priori. Indeed, dynamic and under control deficits of the ships are changeable due to the variable operating conditions on ships and at sea. Based on the given explanations, the most important contributions and novelties of the presented paper are organized as follows:

i) Providing learning machine based on a neural-fuzzy network: combining the fuzzy inference systems, based on logical rules, and the artificial neural network methods that can extract knowledge from numerical information results.

ii) When the performance in GIB systems is enhanced, one witnesses its interference with other controllers' performance. Consequently, in the presented study's system in which there exists a broad multi-variable system, an appropriate control planning would be of fundamental significance. Moreover, the system experiences the best operating conditions, in case of having a genuine coordination between these controls. But, the position of the buoy may be lost or even there may occur continuing conflicts in exercising the controlling signals due to the absence of an appropriate coordination.

iii) Because of the convolution of this system and also the classic controller incapability in diverse working conditions, the study aims at presenting the control parameters with proper adjustments and also an improved neural network.

iv) In this sense, applying the intelligent control law in order to maintain the vessel's healing at the set value and also the position along with having the capability to guarantee high control precision under certain marine environments, can be regarded as the our fundamental aim. Moreover, it is the control law which ought to satisfy the thrust allocation algorithm and propulsion system physical limitation.

## 2. Modeling Buoys

Figure 1 demonstrates the reference coordinate of buoys. In this figure, $OX_0Y_0$ and AXY are the earth-fixed and the body-fixed locations, respectively. If consider symmetrical shape for the ship, the corrugated movement is a combination of volatility and horizontal rotation of ship around the vertical axis. In this model, the climbing mode and the semi-circular vertical and rotational spin are neglected [1]. Thus, the DP of the ship can be formulated as follows:

$$\dot{\eta} = R(\psi)\upsilon, M\dot{\upsilon} + D\upsilon = \tau + \Delta \tag{1}$$

where, $\eta = [x, y, \psi]^\tau$ denotes the earth-fixed locations (x, y) and $\psi$ is the yaw angle with vector type. $\upsilon = [u, r, \upsilon]^\tau$ denotes the moment in surge, yaw and sway presented by means of the thruster system.

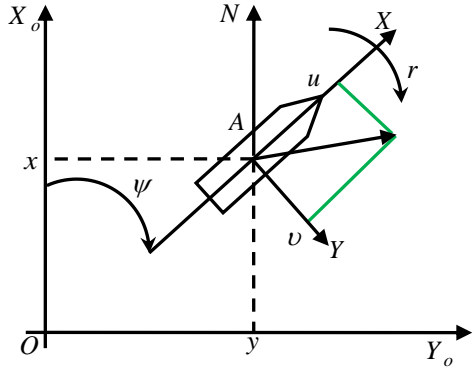

Figure 1. The earth-fixed OX0Y0 and the body-fixed AXY frames of reference

For $\tau = [\tau_1, \tau_2, \tau_3]^\tau$, the control inputs include energy wave movements τ1, oscillatory force τ2, and force τ3. The matrix R (rotation) can be expressed by [2]:

$$R(\psi) = \begin{bmatrix} \cos\psi & -\sin\psi & 0 \\ \sin\psi & \cos\psi & 0 \\ 0 & 0 & 1 \end{bmatrix} \qquad (2)$$

This matrix is orthogonal, that is $R^{-1}(\psi) = R^T(\psi)$, and has the property |R|=1. ‖ is a determinant of the matrix, M is invertible symmetry, and D is a linear damp matrix, respectively.

$$M = \begin{bmatrix} m - X_{\dot{u}} & 0 & 0 \\ 0 & m - Y_{\dot{u}} & -Y_{\dot{r}} \\ 0 & -N_{\dot{v}} & I_z - N_{\dot{r}} \end{bmatrix} \qquad (3)$$

$$D = \begin{bmatrix} X_{\dot{u}} & 0 & 0 \\ 0 & m - Y_v & -Y_r \\ 0 & -N_v & N_r \end{bmatrix} \qquad (4)$$

where Iz, M and m are the inertia factor on the z-axis, a positive constant which used for the standard state of DP and the vessel mass, respectively. The interested reader can refer [1-2] to get more details. Foreign forces and torque, based on the wind, waves, and marine currents, are added in a fixed format. Furthermore, we have the following assumptions in this paper:

1. Both position $\eta = [x, y, \psi]^\tau$ and velocity $\upsilon = [u, \upsilon, r]^\tau$ measurements of the positioning system of ships are available.

2. A) M, D matrices are bounded and unknown. B) $\Delta(t)$ unknown variables are bounded. Moreover, $\bar{\Delta} = [\bar{\Delta}_1, \bar{\Delta}_2, \bar{\Delta}_3]^T$ is an unknown positive vector so that

$$|\Delta_i(t)| \leq \bar{\Delta}_i, i = 1, 2, 3 \tag{5}$$

Note 1. Both conditions of operating buoys and ocean are changing. Therefore, the energy available in the system is always limited in its nature. As a result, the dynamic parameter matrices of M, D of the mathematical model positioning disorders are dynamic and are considered to be variable and limited in the presented study. At this point, in the presented study, the awareness of the bounds of parameters and disorders is not necessary, while it is only the awareness of their existence which is required for analysis purposes [2,3]. Based on mentioned assumptions, the objective is to get the control input data in order to impose the location of (x, y) and $\psi$ known as the ship orientation angle to meet to the preferred objective location of $\eta_d = [x_d, y_d, \psi_d]^\tau$ while lead to minimum voluntary error. The aim is to ensure that the closed-loop DP model signals of all ships are ultimately and uniformly bounded.

## 3. Neural Network

Artificial NN are patterns of information processing created using the imitating biological NN like that of the person brain. The artificial neural network is a data processing system which has been inspired by human brain and has entrusted the processing of data to small, very large processors that work in a networked and parallel manner to solve a problem [9]. Generally, theses networks can be summarized as three main parts:

1. Input layer: they are directly connected with the input data.

2. Hidden layers: The weights among input and hidden units are determined when a hidden unit is enabled.

3. Output Layers: The performance of the output unit depends on the activity of the hidden unit and their connecting coefficients.

Shorting speaking, Figure 2 shows Preston systems used for the improved neural network. The units of each layer may distribute similar entries among themselves while they are not connected to each other. Generally, the units of the input layer are to transfer the input patterns to the rest of the network without doing any processing [9]. Information is processed by the units of hidden layers and output layers. Figure 3 demonstrates the architecture of the three-layer feed-forward neural network. See reference [9] for more information.

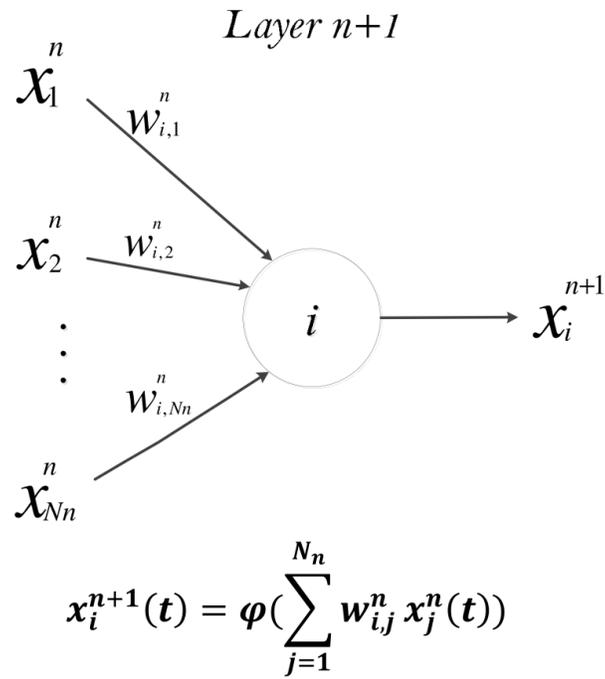

$$x_i^{n+1}(t) = \varphi\left(\sum_{j=1}^{N_n} w_{i,j}^n x_j^n(t)\right)$$

Figure 2. Improved Preston Network

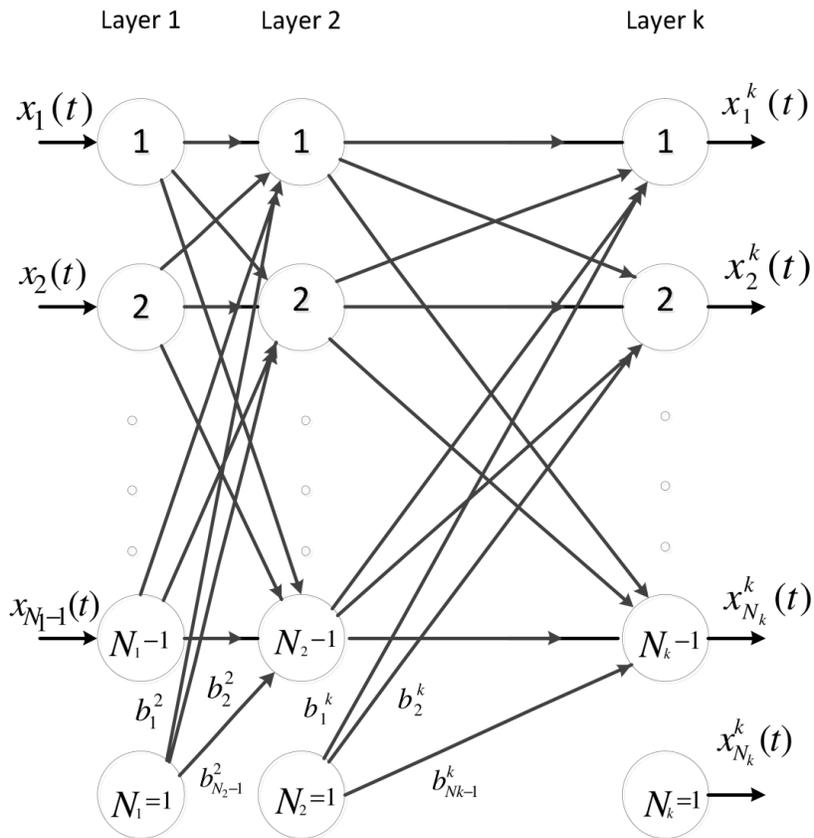

Figure 3. An example view of three-layer feed-forward NN with one hidden layer

## 4. Adaptive Neuro-Fuzzy Network

Generally, fuzzy systems have been applied specifically when we are not able to define or there is little hope for defining crisp rules which are capable of describing the suggested process or system. Consequently, allowing to describe fuzzy rules, whose aim is to fit the explanation of real processes to a larger scale, can be regarded as the most significant advantages of fuzzy systems. Moreover, one can mention the interpretability as the other important advantage of fuzzy systems. In other words, one can point to the possibility of explaining the reason for the appearance of a particular value at the output of a fuzzy scheme. However, in fuzzy systems some specific points should be noticed in order to explain the fuzzy rules, instructions would be necessary, and also the process of fuzzy scheme parameters tuning needs a considerable time specifically when various fuzzy rules are present in the system.

The mentioned disadvantages can be linked to the impossibility of training the fuzzy systems. Note that it is in the NN field that a diametrically opposite situation would be present. Neural networks can be trained by a user despite the fact that it is very hard to apply a priori knowledge regarding the given system and also it is hardly possible to explain the neural system behavior in specific situations. The neural networks have been combined with fuzzy systems by some researchers in an attempt to compensate the advantages of one system with disadvantages of the other. A hybrid system which is referred to as ANFIS (Adaptive-Network Based Fuzzy Inference System) has been suggested. The learning capabilities of ANN can be grasped by the fuzzy-logic based paradigm (ANFIS) in order to amplify the performance of the intelligent system applying priori knowledge. A fuzzy inference system (FIS) can be made by ANFIS applying a given input/output data set. In this sense, membership function parameters of (FIS) can be adjusted by either a single back-propagation method or when combined with a least squares model of method. Consequently, the fuzzy systems will be allowed to learn from the data they are modeling.

The combination of NN and fuzzy logic to extract the best relationship between input and output data leads to increased efficiency of neural network structure training (or trained) on the data received from GPS signals. In other words, the integration of fuzzy logic will increase the flexibility and efficiency of the neural network in learning and extracting definite answers and will avoid complex calculations. Figure 4 indicates neuro-fuzzy modeling. In point of fact, a fuzzy neuron has n weighted inputs of ($w_i$, $x_i$, $i=1, ..., N$) and M outputs. All inputs and weights are real values. In addition, the outputs are positive real values in the range of (0,1), representing a membership value in the fuzzy concept. In other words, the system presents the ownership for a sample inputs.

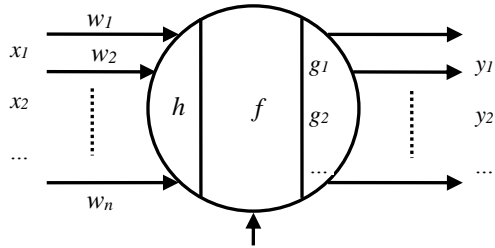

Figure 4. Neuro-fuzzy structure

Relations governing the neuron-fuzzy system take into account the social function of h with the net input of Z, network output of g, and activation function of f, which are all modeled as follows:

$$\begin{aligned} Z &= h[\omega_1 x_1, ..., \omega_n x_n] \\ S &= f[Z - t] \\ Y_j &= g[S] \quad for \quad j = 1, ..., M \end{aligned} \quad (6)$$

If the output for each layer is $O_i^l$ (the i-th node of l), the structure of the model consists of five layers as follows:

Layer 1: after the input data were entered into the system in the form of (y, x) matrix, then, in this layer which uses membership functions, the fuzzy operation is performed. The membership function employed in the present paper to extract fuzzy sets is expressed as follows:

$$\mu_{A_i} = \frac{1}{1+\left(\frac{x-c_i}{a_i}\right)^{2b_i}} \quad (7)$$

where, x denote the input node i, and $s_1 = \{a_i, b_i, c_i\}$ denote the parameter set.

Layer 2, rule nodes: each node denotes the fire strength of the rule.

Layer 3, rule nodes: the ith node calculates by:

$$o_i^3 = \bar{\omega}_i = \frac{\omega_i}{\sum_i^n \omega_i} \quad (8)$$

$\bar{w}_i$ : degree of normalized activity of the i-th rule

Layer 4, result nodes: in the outer layer, each node is equal to:

$$o_i^3 = \bar{\omega}_i f_i = \bar{\omega}_i (p_i x_1 + q_i x_2 + r_i) \quad (9)$$

where, $S_2 = \{p_i, q_i, r_i\}$ is the parameter set for this node.

Layer 5, output nodes: in this layer, each node calculates the final output value as follows (the number of nodes is equal to the number of outputs):

$$o_i^5 = \sum_{i=1}^n \bar{\omega}_i f_i \quad (10)$$

In the neuro-fuzzy model, simulation is done correctly when comparative parameter S1 and subsequent parameter S2 are estimated to be the training and testing values to minimize the model error. In control engineering, fuzzy neural-adaptive networks are applied to model nonlinear functions due to their ability to estimate functions. In this study, in the NN technique for back-stepping, $W^T G(X)$ is the input of $X \in R^m$. $X \in R^m$, $G(X): R^m \to R^l$ and $l > 1$ are the weight constant vector, basic function to extract the nonlinear pattern and the number of nodes in the NN, respectively. $f(X):\Omega_X \to R$; $X = [x_1, x_2, ..., x_m]^T \in \Omega_X$ where $\Omega_X$ is a continuous set in $R^m$. According to the above-mentioned function [9]:

$$f(X) = W^T G(X) + \varepsilon(X), \forall X \in \Omega_X \tag{11}$$

where, $\varepsilon(X)$ is the error of approximation. The basic function is selected as a vector of $G(X) = [g_1(X), g_2(X), ..., g_j(X), ..., g_l(X)]^T$. The Gaussian functions are:

$$g_j(X) = \frac{1}{\sqrt{2\pi h_j}} \exp\left[-\frac{\|X - k_j\|^2}{2h_j^2}\right], j = 1, 2, ..., l \tag{12}$$

$K_j = [k_{j,1}, g_{j,2}, ..., g_{j,m}]^T$ and $h_j$ refer to the center of the receptive circle and Gaussian function width, respectively. Generally, $W^*$ is defined as the value of W which optimizes $|\varepsilon(X)|$. In point of fact, the perfect setting of these parameters is done in accordance to the minimized output error. As previously mentioned, we focus on the simulation error of positioning vessels [2].

$$W^* = \arg \min_{W^* \in R^2} \{\sup_{X \in \Omega_X} |f(X) - \hat{W}^T G(X)|\} \tag{13}$$

In this equation, a vector of ideal fixed weight is just a dummy value which is essential for logical aims and cannot be applied practically. W employed to estimate $W^*$. If $X \in \Omega_X$ then $W^*$ and $\varepsilon(X)$ will be bounded, and for positive constant values of $\varepsilon^*$ and $W_M$, we have: $|W^*| \leq W_M$ and $\|\varepsilon(X)\| \leq \varepsilon^*$.

4.1. Adaptive neuro-fuzzy network control modeling for ship positioning system

This section discusses the proposed controller for DP problem (1). Let $\eta_d = [x_d, y_d, \psi_d]^\tau$ be the ideal location and direction of ships. For the proposed design follow the below steps:

Step 1: Define $Z_1 = \eta - \eta_d$ which is obtained as follows:

$$\dot{Z}_1 = R(\psi)\upsilon - \dot{\eta}_d = R(\psi)\upsilon \tag{14}$$

where, $\upsilon$ and $\alpha_1$ denote the virtual control input data and the virtual control function:

$$\alpha_1 = -R^{-1}(\psi)K_1 z_1 \tag{15}$$

where, $K_1$ dictates a positive constant. We define $z_2 = \upsilon - \alpha_1$; therefore, (14) should be expressed as follows [2]:

$$\dot{Z}_1 = R(\psi)z_2 - K_1 z_1 \tag{16}$$

Consider the Lyapunov's function candidate V1 as follows:

$$V_1 = \frac{1}{2}z_1^T z_1 \tag{17}$$

The derivative of the time of V1 expressed by equation (16) is as follows:

$$\dot{V}_1 = z_1^T (R(\psi)z_2 - K_1 z_1) \tag{18}$$

Step 2: Use (1) and define $Z_2$. A derivative of $Z_2$ is defined as follows:

$$\dot{Z}_2 = M^{-1}[\tau + \Delta - D\upsilon] - \dot{\alpha}_1 = M^{-1}[\tau + \Delta - D\upsilon - M\dot{\alpha}_1] \tag{19}$$

Consider the Lyapunov's function candidate of $V_2$ as follows:

$$V_2 = V_1 + \frac{1}{2}z_2^T M z_2 \tag{20}$$

According to equations (18) and (19), the derivative of the time of V2 is as follows:

$$\begin{aligned}\dot{V}_2 &= \dot{V}_1 + z_2^T M \dot{z}_2 = z_1^T R(\psi)z_2 - z_1^T K_1 z_1 + z_2^T M M^{-1}[\tau + \Delta - D\upsilon \\ &- M\dot{\alpha}_1] \leq z_1^T R(\psi)z_2 - z_1^T K_1 z_1 + z_2^T [\tau + \text{sgn}(z_2)\overline{\Delta} - D\upsilon - M\dot{\alpha}_1], \\ &\text{where}, \text{sgn}(z_2) = \text{diag}([\text{sgn}(z_{2,1}), \text{sgn}(z_{2,2}), \text{sgn}(z_{2,3})])\end{aligned} \tag{21}$$

where, parameters D, M and $\overline{\Delta}$ will be unknown [1–6]. The proposed enhanced NN to estimate and adjust them by:

$$\Theta^{*T} S(Z) + E(Z) = D\upsilon + M\dot{\alpha}_1 - \text{sgn}(z_2)\overline{\Delta} \tag{22}$$

where, $Z = [\eta^T, \upsilon^T, \dot{\alpha}_1^T]^T$ denote the input vector of the proposed neural network, and $S(Z) = [S_1^T(Z), S_2^T(Z), S_3^T(Z)]^T$ is the basic function vector.

$$\Theta^* = \begin{bmatrix} \theta_1^{*T} & 0_{1\times l} & 0_{1\times l} \\ 0_{1\times l} & \theta_2^{*T} & 0_{1\times l} \\ 0_{1\times l} & 0_{1\times l} & \theta_3^{*T} \end{bmatrix}_{3\times 3l} \tag{23}$$

The constant weight matrix is ideal with $\theta_i^* = [\theta_{i,1}^*, \theta_{i,2}^*, \theta_{i,l}^*]^T$, (i = 1, 2, 3), and E(Z) approximation error vector is obtained from unknown dynamic and floating disturbances. (22) is written as follows:

$$\dot{V}_2 \leq Z_1^T R(\psi) Z_2 - Z_1^T K_1 z_1 + z_2^T [\tau - \Theta^{*T} S(Z) - E(Z)] \tag{24}$$

By considering $\hat{\theta}_i, i = 1, 2, 3$ from approximation of $\theta_i^*, i = 1, 2, 3$, the law control is designed as follows:

$$\tau = -R^T(\psi) z_1 - K_2 z_2 + \hat{\Theta}^T S(Z) \tag{25}$$

The updated parameters of the law control are defined as follows:

$$\dot{\hat{\theta}}_i = \Gamma_i [S_i(Z) z_{2,i} + \sigma_i \hat{\theta}_i], i = 1, 2, 3 \tag{26}$$

where, $K_2 \in R^{3\times 3}$ is defined as a fixed positive matrix by taking into account the disturbances as follows:

$$\hat{\Theta} = \begin{bmatrix} \tilde{\theta}_1^T & 0_{1\times l} & 0_{1\times l} \\ 0_{1\times l} & \tilde{\theta}_2^T & 0_{1\times l} \\ 0_{1\times l} & 0_{1\times l} & \tilde{\theta}_3^T \end{bmatrix}_{3\times 3l} \tag{27}$$

$\Gamma_i = \Gamma_i^T \in R^{l\times l}, \sigma_i > 0, (i = 1, 2, 3)$ is defined as the positive design matrix and is fixed. In addition, consider the candidate $V_{2a}$ of the Lyapunov's additional function [2]:

$$V_{2a} = V_2 + \frac{1}{2} \tilde{\theta}_i^T \Gamma_i^{-1} \hat{\theta}_i \tag{28}$$

$\tilde{\theta}_i = \hat{\theta}_i - \theta_i^*, i = 1, 2, 3$ is an approximate error of weight vector. According to equation (26), the derivative of Lyapunov's function (28) is as follows:

$$\dot{V}_{2a} = \dot{V}_2 + \sum_{i=1}^{3} \tilde{\theta}_i^T \Gamma_i^{-1} \dot{\tilde{\theta}}_i \leq Z_1^T R(\psi) Z_2 - Z_1^T K_1 z_1 + z_2^T [\tau - \Theta^{*T} S(Z) - E(Z)] + \sum_{i=1}^{3} \tilde{\theta}_i^T \Gamma_i^{-1} \dot{\tilde{\theta}}_i \quad (29)$$

With a direct placement of (26) and (27)-(29), we have:

$$\dot{V}_{2a} \leq -z_1^T K_1 z_1 - z_2^T K_2 z_2 + z_2^T \tilde{\Theta}^T S(Z) - z_2^T E(Z) - \sum_{i=1}^{3} \tilde{\theta}_i^T [S_i(Z) z_{2,i} + \sigma_i \tilde{\theta}_i]$$

$$\leq -z_1^T K_1 z_1 - z_2^T K_2 z_2 - z_2^T E(Z) - \sum_{i=1}^{3} \sigma_i \tilde{\theta}_i^T \tilde{\theta}_i \quad (30)$$

where, $\tilde{\Theta} = \hat{\Theta} - \Theta^*$ is the estimation of the error of the weight matrix. On completion of the squares and inequalities, we have:

$$-z_2^T E(Z) \leq \frac{z_2^T z_2}{2} + \frac{\|E^*\|^2}{2} \quad (31)$$

$$-\sigma_i \tilde{\theta}_i^T \hat{\theta}_i \leq -\frac{\sigma_i \tilde{\theta}_i^T \tilde{\theta}_i}{2} + \frac{\sigma_i \|\theta_i^*\|^2}{2} \leq -\frac{\sigma_i \tilde{\theta}_i^T \tilde{\theta}_i}{2} + \frac{\sigma_i \theta_{i,M}^2}{2} \quad (32)$$

where, $E^*$ is the revival error vector bound, and $\theta_{i,M}^2$ is the ideal weight revival error vector bound of $\theta_i^*$ in Hypothesis 3. According to the above-mentioned formulations, the overall plan for system design is summarized as follows:

1. As Hypotheses 1-3 show, robust adaptive control law followed by equations (2) and (21) for nonlinear DP model (1) of floating buoys have been developed in accordance to the dynamic and uncertain disturbances. The control law ensures that the location (x, y) and the real orientation ψ are converged in the desired target position and orientation [2]. By selecting appropriate design parameters $\sigma_i, i = 1, 2, 3$, the positive design factors matrix $\Gamma_i > 0, i = 1, 2, 3$ and K₁ and K₂ with suitable dimensions are determined by the proposed algorithm; by solving (30), we have:

$$0 \leq V_{2a}(t) \leq \frac{c}{\Phi} + [V_{2a}(0) - \frac{c}{\Phi}]e^{-\Phi t}$$

$$\Phi = \min\{2\lambda_{\min}(K_1), 2\lambda_{\min}\{(K_2 - \frac{I_{3\times 3}}{2})M^{-1}\}, \beta\} > 0 \qquad (33)$$

$$c = 0.5 \|E^*\|^2 + \sum_{i=1}^{3} \frac{\sigma_i \theta_{i,M}^2}{2} > 0, K_2 > \frac{I_{3\times 3}}{2}$$

As can be observed in equation (33), $V_{2a}(t)$ is ultimately bounded uniformly. As a result, signals $\tilde{\theta}_i, i = 1,2,3; z_1, z_2$ are ultimately bounded uniformly. Since $\eta_d$ is bounded, it can be concluded that $\eta$ is also bounded. Based on the Hypothesis 3, the coefficients of $\theta_i^*, i = 1,2,3$ are bounded and, as a result, $\hat{\theta}_i, i = 1,2,3$ are also bounded. By Eqs. (33) and (28), one gets:

$$\|z_1\| \leq \sqrt{\frac{2c}{\Phi} + 2[V_{2a}(0) - \frac{c}{\Phi}]e^{-\Phi t}} \qquad (34)$$

This equation shows that for each $\mu_{z_1} \leq \sqrt{2c/\Phi}$, there is a fixed value of $T_{z_1} > 0$; thus $\|z_1\| \leq \mu_{z_1}$ can be defined for all $T_{z_1} < t$. When $\sqrt{2c/\Phi}$ can become small to the extent that is required through K1 and K2 with the appropriate selection of design parameters $\sigma_i, \Gamma_i, i = 1,2,3$, the positioning error can increase accuracy. Therefore, the position and direaction of the ship can be converged $\eta_d = [x_d, y_d, \psi_d]^\tau$ and can also be adequately maintained in $\eta_d$. Figure 5 presentes the flowchart of the proposed technique.

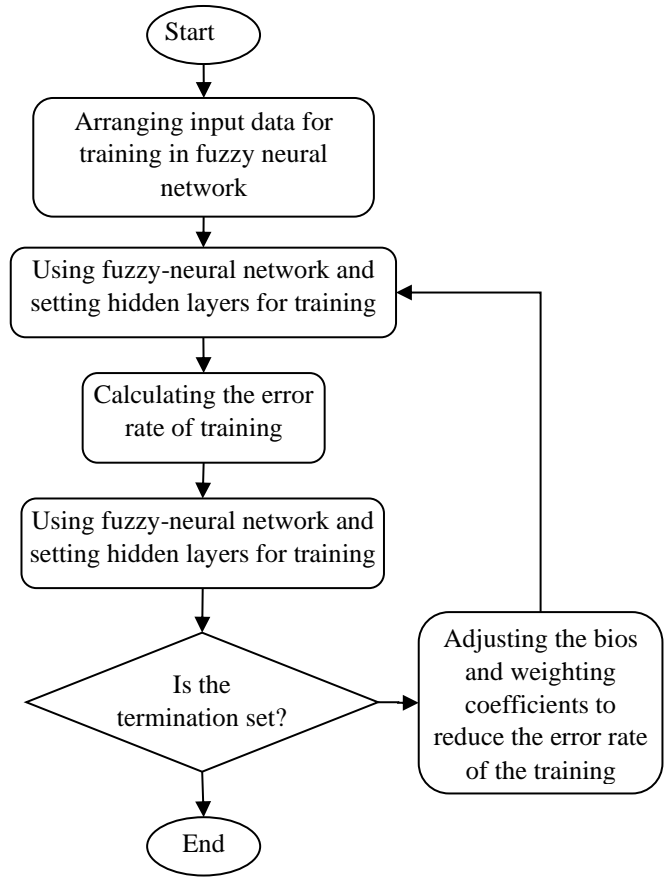

Figure 5 - Flowchart of the proposed algorithm

Maintaining the stability, when the system enters the saturation phase, has been regarded as the major challenge by means of a linear control scheme subject to input saturation [2,21]. In this sense, the induced $L_2$ norm or the finite $L_2$ gain of an operator H can be defined as $\|H\|_{i,2} = \sup_{x \in L_2} \frac{\|Hx\|_2}{\|x\|_2}$ where $\|x\|_2 = \sqrt{\int_0^\infty \|x\|^2}$ denotes the $L_2$ norm of the vector x(t), and $\|x\|$ is its Euclidean norm. In this regard, $\|x\|_2$ if $x \in L_2$ is finite. Consequently, the weighted $L_2$ norm can be defined as

$$\|x_t\|_{2\delta} = \sqrt{\int_0^t \exp(-\delta(t-\tau)) x^T(\tau) d\tau} \ldots\ldots\ldots \qquad . \quad (35)$$

Where $\delta \geq 0$ is a constant. Likely, if $\|x_t\|_{2\delta}$ is finite, $x \in 2\delta$. $L_\infty$ can propose a definition for $\|x_t\|_\infty = \sup_{t \geq 0} |x(t)|$ Where $x \in L_\infty$ If $\|x\|_\infty$ is present. Grimm et al. [22] have presented a decentralized saturation function for a vector $u \in R^{n_u}, n_u \geq 1$, but, herein, we concentrate on scalars. Accordingly, the signals can be bounded in the closed-loop and also the reference signal can be closely tracked by the plant output. The added nonlinearities, i.e. actuator saturation constraints, will bring about more challenges in adaptive systems. It is also worth mentioning that the ad-hoc approach stops the tracking error which has been induced using the actuator constraints. Yet, the closed-loop system's stability would not be established. In this sense, we aimed at applying an indirect adaptive control structure in which we attempted to establish the closed-loop system stability unaccompanied by stopping the adaptation. In case of adding and subtracting $\tilde{A}e$ and $\tilde{B}sat(u_{act})$ to the tracking error dynamics:

$$\dot{e} = \hat{A} + \hat{B}sat(u_{act}) + Ax_r - \tilde{A}e - \tilde{B}sat(u_{act}) \tag{36}$$

Where $-\tilde{A}e$ and $-\tilde{B}sat(u_{act})$ can be regarded as the terms because of the parameter errors. Expressing the saturated plant input $sat(u_{act})$, the state vector X and also the tracking error $E = [E_1, E_2, ..., E_n]^T$, requires $Em^2$ to be written as:

$$Em^2 = -\tilde{A}\frac{1}{s+\lambda}x - \tilde{B}\frac{1}{s+\lambda}sat(u_{act}) = -\tilde{A}\frac{1}{s+\lambda}(e + x_r) - \tilde{B}\frac{1}{s+\lambda}sat(u_{act}) \tag{37}$$

Operating on each side of above equation with $(s+\lambda)$ resulted in obtaining the following:

$$(s+\lambda)Em^2 = -\tilde{A}e - \tilde{B}sat(u_{act}) - \dot{\tilde{A}}\frac{1}{s+\lambda}e - \dot{\tilde{B}}\frac{1}{s+\lambda}sat(u_{act}) - \tilde{A}x_r - \dot{\tilde{A}}\frac{1}{s+\lambda}x_r \tag{38}$$

Moreover, we can also combine it with Eq. (38):

$$\dot{e} = \hat{A}e + (s+\lambda)Em^2 + \dot{\tilde{A}}\frac{1}{s+\lambda}e + \dot{\tilde{B}}\frac{1}{s+\lambda}sat(u_{act}) + \tilde{A}x_r + \dot{\tilde{A}}\frac{1}{s+\lambda}x_r + \hat{B}sat(u_{act}) + Ax_r \tag{39}$$

In this regard, we can define a new vector ($\bar{e} = e - Em^2$) which can be substituted into the Eq. (39):

$$\dot{\bar{e}} = \hat{A}\bar{e} + (\lambda I + \hat{A})Em^2 m + \hat{A}\frac{1}{s+\lambda}e + \hat{B}\frac{1}{s+\lambda}sat(u_{act}) + \dot{A}x_r + \hat{A}\frac{1}{s+\lambda}x_r + \hat{B}sat(u_{act}) \quad (40)$$

$det(sI - \hat{A}(t))$ roots have to be in negative half-plane, if we aim at establishing the stability of the homogeneous part of the Eq (40). Setting the upper projection limit for A as negative and, then, properly initializing the parameter estimation can satisfy $A \in R$ in the first order plants. It is worth mentioning that for higher order systems guaranteeing the property by projection is not trivial. Consequently, modifying the projection operator can allow $\hat{A}$ to possess eigenvalues with negative real parts. In this sense, this will happen only when the plant is stable. The homogeneous part of Eq (40) cannot be affected by the modified linear quadratic controller output $u$ or $u_{act}$, specifically when we have the saturation phase and the control input is $sat(u_{tac})$. Moreover, it can be illustrated as a constant equal to $u_{act}$ or $-u_{act}$. Although the projection can bound A and B, definition can bound $sat(u_{tac})$ and the reference signal $x_r$. $Em, \dot{\hat{A}}, \dot{\hat{B}} \in L_\infty$ is guaranteed by the adaptive law.

## 5. Numerical Simulation and Analysis

In this section, numerical simulation was performed on a buoy so as to check the effects of robust adaptive control law based on the proposed algorithm. The information about this buoy is available in reference [2]. Dynamic model parameters in (1) are provided as follows:

$$M = \begin{bmatrix} 5.3122 \times 10^6 & 0 & 0 \\ 0 & 8.2831 \times 10^6 & 0 \\ 0 & 0 & 3.7454 \times 10^9 \end{bmatrix}, D = \begin{bmatrix} 5.0242 \times 10^4 & 0 & 0 \\ 0 & 2.7229 \times 10^5 & -4.3933 \times 10^6 \\ 0 & -4.3933 \times 10^6 & 4.1894 \times 10^8 \end{bmatrix} \quad (41)$$

In the simulations, two types of disturbances are considered. The first type is fixed disturbances $\Delta = [1000N, 2000N, 1500Nm]^\tau$ and the second one is the biasing force and torque vector $b \in R^3$,

$b = -T^{-1}b + \psi_n$, $\Delta = R^\tau(\psi)b$. The Gaussian white noise vector with zero mean, constant-time bias $\psi = \text{diag}[1000,1000,1000]$, and $T = \text{diag}[1000,1000,1000]$ dictates a diagonal scale matrix with the range of n, which is explained in accordance with the information in [9]. Moreover, they have copied the disrupting environmental impacts by including second-order waves, ocean currents, and wind. With regard to the optimum position and orientation of the ship $\eta_d = [0m, 0m, 0°]^T$, primary modes are selected as $\eta(0) = [10m, 10m, 10°]^T$. First, the PID control scheme is studied for the simulation. PID control parameters $\tau_{PID} = K_p \tilde{\eta} + K_i \int_0^t \tilde{\eta}(\tau)d\tau + K_d (d\tilde{\eta}(t)/dt)$ are calculated as follows:

$$K_p = \begin{bmatrix} 3000 & 0 & 0 \\ 0 & 9000 & 0 \\ 0 & 0 & 10^8 \end{bmatrix}, K_i = \begin{bmatrix} 5 & 0 & 0 \\ 0 & 50 & 0 \\ 0 & 0 & 30 \end{bmatrix}, K_d = \begin{bmatrix} 5\times10^4 & 0 & 0 \\ 0 & 7\times10^4 & 0 \\ 0 & 0 & 300 \end{bmatrix} \quad (42)$$

Using the initial conditions and parameters of the controller PID, we have put the system under various stresses. Figures 6 and 7 show the simulation results. Albeit, the PID controller shows acceptable performance, but, its performance in the presence of time-varying disturbance is not acceptable.

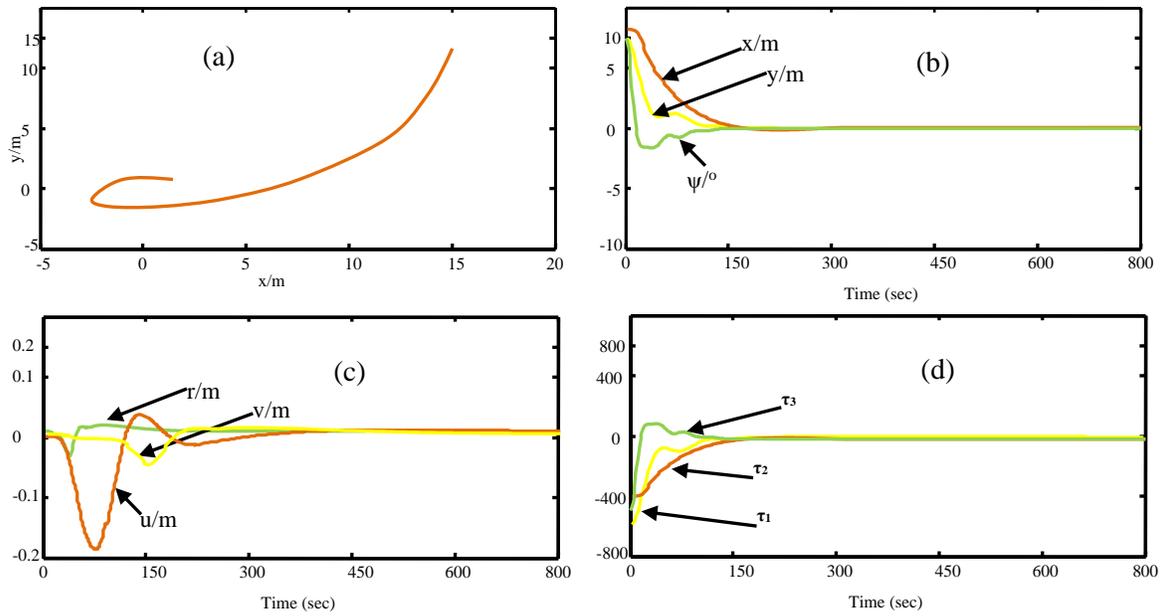

Figure 6- (a) Curved path of buoy in the xy plane with PID controller in constant disturbance, (b) Actual position curves (x, y) and ψ orientation to time with PID controller in constant disturbance, (c) Initial speed of wave motion curves u, initial speed of oscillation v, and horizontal rotation speed of ship around the vertical axis with respect to time with PID controller in constant disturbance, and (d) Motion control force curves τ1, damping control power τ2, and horizontal torque of ship to vertical axis τ3 with respect to time and with PID controller in constant disturbance

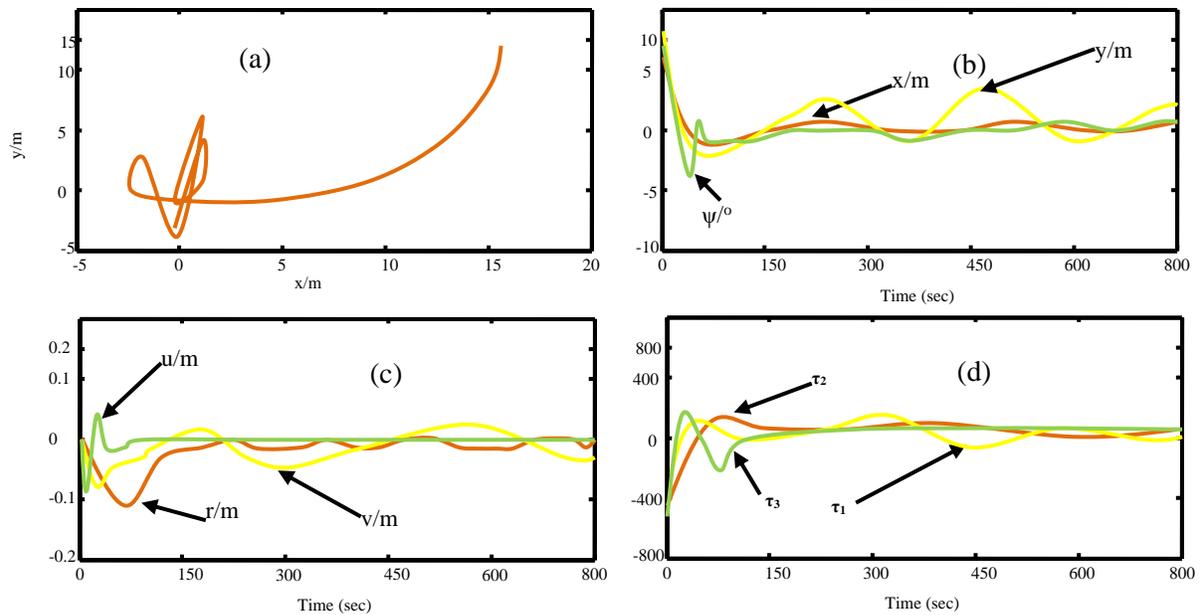

Figure 7- (a) Curved path of buoy in the XY plane with PID controller in variable disturbance, (b) Actual position curves (x, y) and ψ orientation to time with PID controller in variable disturbance, (c) Initial speed of wave motion curves u, the initial speed of oscillation v, and horizontal rotation speed of ship around the vertical axis with respect to time with PID controller in variable disturbance, and (d) Motion control force curves τ1, damping control power τ2, and horizontal torque of ship to vertical axis τ3 with respect to time and with PID controller in variable constant disturbance

In what follows, the proposed controller based on the improved NN is studied by simulation. As it is determined in these simulations, the same initial states and optimal positions and orientations are acceptable. A number of network nodes are $l = 3^9$ in total which are equally placed with their centers in coordinates of $\begin{Bmatrix} [-2,10] \times [0,10] \times [-0.05, 0.2] \times [-0.35, 0.05] \\ \times [-0.012, 0.004] \times [-0.5, 0.2] \times [-0.7, 0] \times \\ [-0.14, 0.04] \end{Bmatrix}$. Additionally, the width of the Gaussian function $h_{ij} = 1 (i = 1,2,3; j = 1,2,3)$ with preliminary estimates of random values $\hat{\theta}_i(0) = [\hat{\theta}_{i,1}(0), ..., \hat{\theta}_{i,l}(0)]^T$ is selected randomly in [0,1]. The employed design matrices can be defined by:

$$K_1 = \begin{bmatrix} 0.037 & 0 & 0 \\ 0 & 0.063 & 0 \\ 0 & 0 & 0.832 \end{bmatrix}, K_2 = \begin{bmatrix} 5 \times 10^4 & 0 & 0 \\ 0 & 6 \times 10^4 & 0 \\ 0 & 0 & 5.4 \times 10^4 \end{bmatrix}$$
$$\Gamma_1 = \Gamma_2 = \Gamma_3 = diag[0.1] \in R^{3^9 \times 3^9},$$
$$\sigma_1 = \sigma_2 = 2.13, \sigma_3 = 0.302$$
(43)

Figure 8.a demonstrates that the proposed controller is capable of tolerating the ship which is converged to the desired point. Figure 8.b indicates the position and angle of deviation from the ship, showing that the true position of the ship (x, y) and the actual orientation ψ can approach the appropriate position and orientation $\eta_d = [0m, 0m, 0^\circ]$ in approximately 80 seconds. Figure 8.c shows the early speed, initial velocity fluctuations, and deviations from the path of ships. Figure

8d indicates the torque and the corresponding control, and smooth control inputs are acceptable in this figure. As observed in Figure 8e, the average weight $\|\hat{\theta}_1\|, \|\hat{\theta}_2\|, \|\hat{\theta}_3\|$ of the error vector time is bounded. It can be found the proposed model of all signals of the closed-loop DP system of the ship is finally bounded uniformly, as it was proven in Theory 1. The obtained results of the variable disturbance plotted in Figure 9 show the same control functions in Figure 6 with the constant disturbance. The robust adaptive control scheme based on FNN target is acceptable in both constant and variable disturbing conditions. The strong resistance of adaptive controllers to disturbance is evident.

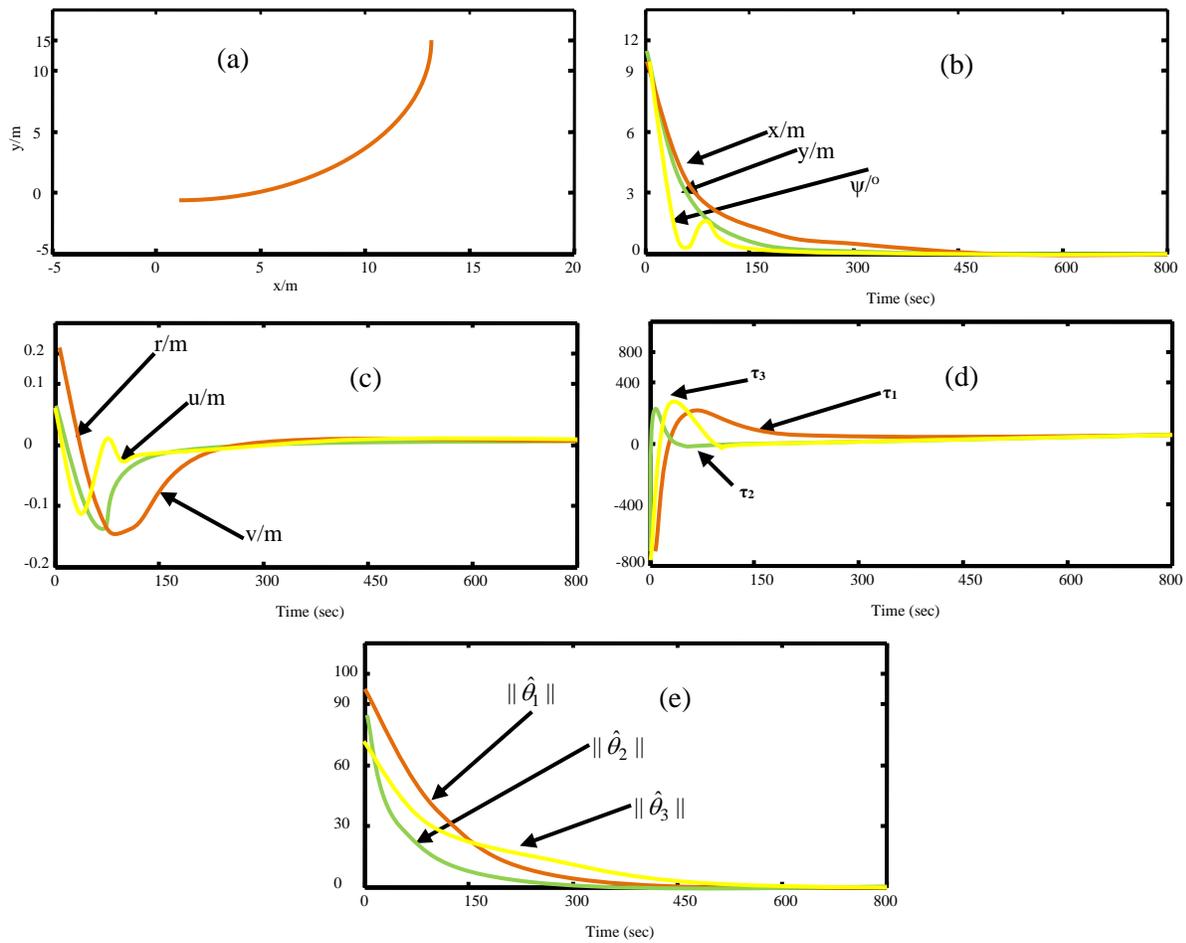

Figure 8- (a) Curved path of buoy in the XY plane with our controller in constant disturbance, (b) Actual position curves (x, y), ψ orientation to time with proposed controller in constant disturbance, (c) Initial speed of wave motion curves u, initial speed of oscillation v, and horizontal rotation speed of ship around the vertical axis with respect to time with proposed controller in constant disturbance, and (d) Motion control force curves τ1, damping control power τ2, horizontal torque of ship to vertical axis τ3 with respect to time and with proposed controller in constant disturbance, and (e) Curves of $\|\hat{\theta}_1\|, \|\hat{\theta}_2\|, \|\hat{\theta}_3\|$ with respect to time with proposed controller in constant disturbance

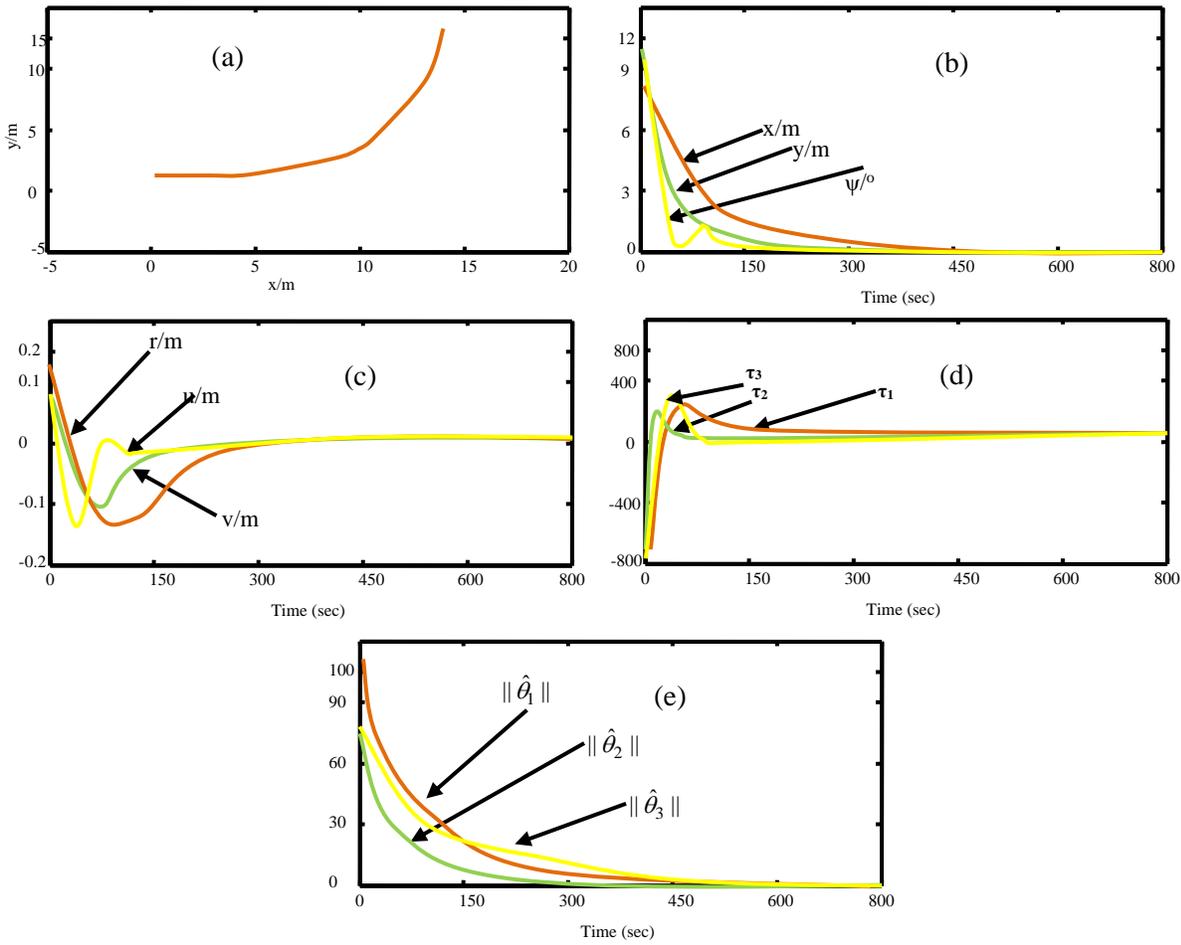

Figure 9- (a) Curved path of buoy in the XY plane with our controller in variable disturbance, (b) Actual position curves (x, y), ψ orientation to time with proposed controller in variable disturbance, (c) Initial speed of wave motion curves u, initial speed of oscillation v, and horizontal rotation speed of ship around the vertical axis with respect to time with proposed controller in variable disturbance, (d) Motion control force curves τ1, damping control power τ2, and horizontal torque of ship to the vertical axis τ3 with respect to time and with proposed controller variable

constant disturbance, and (e) Curves of $\|\hat{\theta}_1\|, \|\hat{\theta}_2\|, \|\hat{\theta}_3\|$ with respect to time with proposed controller in variable disturbance

To compare the proposed controller with other existing controllers, Figure 10 presents the simulation results of our model with only the NN.

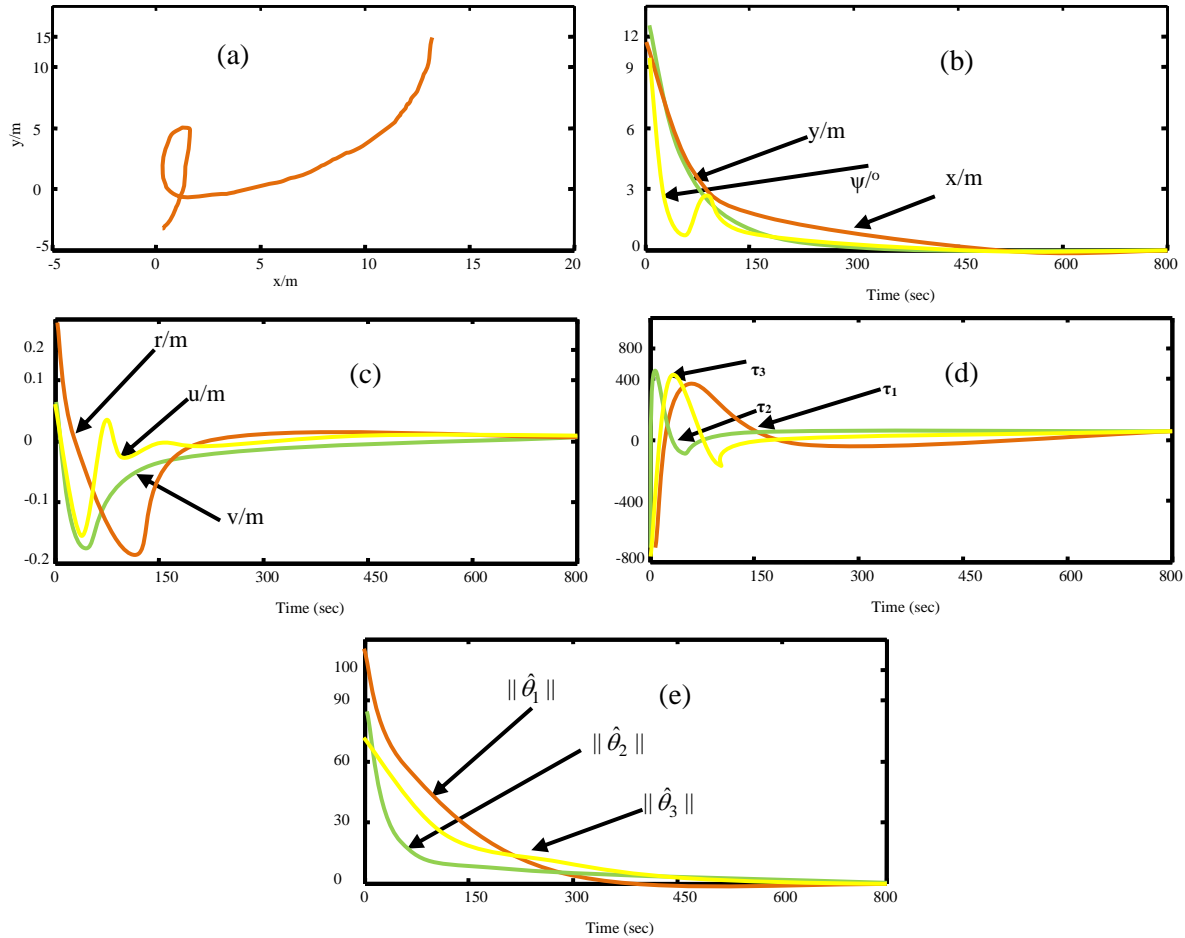

Figure 10- (a) Curved path of buoy in the XY plane with NN controller in constant disturbance, (b) Actual position curves (x, y), ψ orientation to time with NN controller in constant disturbance, (c) Initial speed of wave motion curves u, initial speed of oscillation v, and horizontal rotation speed of ship around the vertical axis with respect to time with NN controller in constant disturbance, and (d) Motion control force curves τ1, damping control power τ2, horizontal torque of ship to vertical axis τ3 with respect to time and with NN controller in constant disturbance, and (e) Curves of $\|\hat{\theta}_1\|, \|\hat{\theta}_2\|, \|\hat{\theta}_3\|$ with respect to time with NN controller in constant disturbance

Despite the fact that NN controller performances are not good under time-variant conditions, it is in Figure 10 that we witness a satisfactory performance of NN scheme under constant disturbances.

**6-Conclusion**

Governments and industrial organizations around the world are cooperating to increase the efficiency standards for electronic chart display and information system employed for positioning electronic devices. This system has made a breakthrough in maritime navigation and plays a leading role in replacing paper and paper maps. Positioning systems have a crucial role in the management of the existing equipment at ports and docks. In this sense, excellent maneuverability, ease of changing the position, and positioning independency to water depth and its avoidance of destroying the seabed can be regarded as the most significant advantages of the dynamic positioning. The technique of dynamic positioning has been utilized in a variety of ways, i.e. offshore exploration, drilling, pipe-laying etc., and researchers have consequently paid considerable attentions to it regarding the fact that there has been a continuous growth in the development and employment of oceans. In theory, with the help of Lyapunov's function, it was proven that adaptive control strategy developed by our proposed algorithm can help buoys reach the proposed target position and warranty the uniformly bounded signals of the closed-loop model. This study indicated that the control model requires no initial knowledge about the dynamics and disturbances of buoys and it is done through the basic pivot function networks which are considered as approximated function in back-stepping controller design. Thus, the dynamic positioning of buoys in different environmental conditions, that are more practical, can be guaranteed. The presented approach's effectiveness and robustness have been validated by the simulation comparisons and results on the dynamic positioning of a supply ship.